\documentstyle[12pt]{article}
\begin{document}
\newcommand{\ds}{\displaystyle}
\newcommand{\be}{\begin{equation}}
\newcommand{\ee}{\end{equation}}
\newcommand{\sgn}{\mbox{sign}}
\newcommand{\gtrsim}{\; \raisebox{-1ex}
   {$ \stackrel{\textstyle >}{\sim}$} \;}
\begin{flushright}
ITP-93-16E \\
hep-th/9311124 \\
March 1993 \\
\vskip 1.0cm
\end{flushright}
\begin{center}
{\Large \bf Distance dependent statistics in a P,T-invariant model}\\
\vspace{1cm}
{\large E.V.~Gorbar\footnote{e-mail: eppaitp@gluk.apc.org},
Stefan V.~Mashkevich\footnote{e-mail: mash@phys.unit.no,
gezin@gluk.apc.org}}\\[.5cm]
{\it Institute for Theoretical Physics, 252143 Kiev, Ukraine}
\end{center}
\vspace{.5cm}
\begin{abstract}
It is shown that in the P,T-invariant model with the mixed
Chern-Simons term the interaction of charge carriers leads to effective
changing of their statistics,  which  depends  on  distance  between
them. In particular, in the limit of large  distances
fermions  effectively turn into bosons.
\end{abstract}
\newpage

The idea of fractional statistics in two dimensions
\cite{r1,r2} and its possible relevance to real physical
systems attract attention from different points of view. The
standard mechanism \cite{r3,r4} leading to appearance
of such statistics is associated with including the
Chern-Simons term in the gauge field action. However,
such term breaks P and T symmetry, which is a serious
obstacle to potential realization  of fractional statistics
in nature. (As is well known, it is only in the systems of
neutral kaons that a weak breakdown of  T  invariance has been
discovered experimentally.) In  essence,  Laughlin's  anyon
ansatz \cite{r5} in the explanation of fractional quantum
Hall effect, where it is the external magnetic field which
breaks P and T, remains the only application of  the  idea
of  fractional  statistics which can be regarded as
confirmed experimentally.

There had been numerous attempts to apply  this  idea
to  construct a theory of high-$T_c$ superconductivity,
starting  from  the works by Laughlin \cite{r6}. However,
such direct attempts apparently  are not relevant, since
the available experimental data do not confirm \cite{r7}
the P and T breakdown effects predicted within the framework
of the anyonic scenario of superconductivity. In connection
with  this, a P and T invariant model of  superconductivity,
which  involves a mixed Chern-Simons term, was put forward
\cite{r8}-\cite{r11}. The authors of \cite{r11} argue that
in this model the charge carriers do not  acquire  fractional
statistics, so that the model is  not  the  one  of  ``anyonic
superconductivity'', despite the presence of the  Chern-Simons
term. In fact, as we will see, this statement holds only  so
far  as  the distance between the charges is kept either much
more or much less than the characteristic scale of interaction
determined by the coupling  constants. In the present work,
we will study  the  case  of  arbitrary distances and show
that the model under consideration gives rise  to effectively
distance dependent statistics, which is in a certain sense
a generalization of fractional statistics \cite{r12,r13};
moreover, it turns out that composites of arbitrary number
of fermions at  large  distances behave as bosons.
We reason  that  the  discovered possibility of having
such statistics in  P  and  T  invariant models makes
it more likely to  appear  in  real physical systems.
We would like to note the difference of
the  case under consideration from the one of Ref.\cite{r14},
where P and T  are  conserved only macroscopically; in the
model at hand, P and  T  invariance is inherent in the
microscopic equations of motion.

The Lagrangian of the model \cite{r11} is
\be \label{e1}
{\cal L} = - \frac{1}{4g^2} f_{\mu\nu} f^{\mu\nu}
+ \bar{\psi} (i \partial \hspace{-0.55em} / \hspace{0.2em}
- \tau_3 a \hspace{-0.5em} / \hspace{0.2em}
- e A \hspace{-0.5em} / \hspace{0.2em}
- \Delta) \psi - \frac{1}{4\sqrt{\partial^2}}
F_{\mu\nu} F^{\mu\nu} +
\frac{\sgn (\Delta) e}{2 \pi} \epsilon^{\mu\nu\rho}
A_{\mu} f_{\nu\rho} .
\ee
Here
\be \label{e2}
f_{\mu\nu} = \partial_{\mu} a_{\nu} - \partial_{\nu} a_{\mu} \; ,
\; F_{\mu\nu} = \partial_{\mu} A_{\nu} - \partial_{\nu} A_{\mu} \; ,
\ee
$A_{\mu}$ is the electromagnetic field and $a_{\mu}$  the
so-called  statistical gauge field; for the coupling
constant $g$  one has $g^2 \sim J ,$  where $J$  is the
parameter of the Hubbard model Hamiltonian. The model
involves two species of fermions unified in a
four-component bispinor
$\psi = \left( \begin{array}{c} \!\! \psi_1 \!\! \\
\!\! \psi_2 \!\! \end{array} \right) .$
The $\gamma$-matrices then form a reducible representation
\be \label{e3}
\gamma_0 = \left( \begin{array}{cc}
 i \sigma_3 & 0 \\ 0 & -i \sigma_3
 \end{array} \right) ,
\gamma_1 = \left( \begin{array}{cc}
 i \sigma_1 & 0 \\ 0 & -i \sigma_1
 \end{array} \right) ,
\gamma_2 = \left( \begin{array}{cc}
 i \sigma_2 & 0 \\ 0 & -i \sigma_2
 \end{array} \right) ,
\ee
and
\be \label{e4}
\tau_3 = \left( \begin{array}{cr}
 I & 0 \\ 0 & -I \end{array} \right) .
\ee
The parity operator acts on the four-component bispinor as
\be \label{e5}
P_4 \psi = \left( \begin{array}{cc}
 0 & P_2 \\ P_2 & 0 \end{array} \right) \psi ,
\ee
so the mass term
$\Delta \bar{\psi} \psi = \Delta \bar{\psi_1} \psi_1 -
\Delta \bar{\psi_2} \psi_2 $ is P-invariant.

The field equations corresponding to the
Lagrangian (\ref{e1}) read
\be \label{e6}
\begin{array}{rcl}
 \frac{1}{g^2} \partial_{\mu} f^{\mu\nu} +
 \frac{se}{2\pi} \epsilon^{\nu\mu\lambda} F_{\mu\lambda} &
 = & j_3^{\nu} , \\  \\
 \frac{1}{\sqrt{\partial^2}} \partial_{\mu} F^{\mu\nu} +
 \frac{se}{2\pi} \epsilon^{\nu\mu\lambda} f_{\mu\lambda} &
 = & j^{\nu} ,  \end{array}
\ee
where
\be \label{e7}
j^{\nu} = \bar{\psi} \gamma^{\nu} \psi ,
j_3^{\nu} = \bar{\psi} \gamma^{\nu} \tau_3 \psi ,
s = \sgn(\Delta) .
\ee
For our purposes it will be
sufficient to restrict ourselves to a purely quantum
mechanical treatment  of the problem. Therefore one
can simply impose the  Lorentz  gauge  conditions
$ \partial_{\mu} A^{\mu} = 0, \partial_{\mu} a^{\mu} = 0 $
so that (\ref{e6}) becomes
\be \label{e8}
\begin{array}{rcl}
 \frac{1}{g^2} \partial^2 a^{\mu} + \frac{se}{\pi}
 \epsilon^{\mu\nu\lambda} \partial_{\nu} A_{\lambda} &
 = & j_3^{\mu} , \\ \\
 \sqrt{\partial^2} A_{\mu} + \frac{se}{\pi}
 \epsilon^{\mu\nu\lambda} \partial_{\nu} a_{\lambda} &
 = & j^{\mu} . \end{array}
\ee
To solve these equations for given
$j^{\mu} , j_3^{\mu} ,$ Fourier transformation
may be applied:
\be \label{e9}
 \begin{array}{rcl}
 - \frac{p^2}{g^2} \tilde{a}^{\mu}(p) + \frac{ise}{\pi}
 \epsilon^{\mu\nu\lambda} p_{\nu} \tilde{A}_{\lambda} (p) &
 = & \tilde{j}_3^{\mu} (p) , \\ \\
 \sqrt{-p^2} \tilde{A}^{\mu}(p) + \frac{ise}{\pi}
 \epsilon^{\mu\nu\lambda} p_{\nu} \tilde{a}_{\lambda} (p) &
 = & \tilde{j}^{\mu} (p) , \end{array}
\ee
and after a straightforward calculation one comes to
\be \label{e10}
 \begin{array}{rcl}
 \tilde{A}^{\mu}(p) & = & \tilde{\cal B}(p) \tilde{j}^{\mu}(p)
 - \frac{s\pi}{e} f \tilde{\cal D}^{\mu\nu}(p)
 \tilde{j}_{3\nu}(p) , \\ \\
 \tilde{a}^{\mu}(p) & = & - \frac{s\pi}{e} f
 \tilde{\cal D}^{\mu\nu}(p) \tilde{j}_{\nu}(p) +
 g^2 \tilde{\cal C}(p) \tilde{j}_3^{\mu}(p) , \end{array}
\ee
where
\be \label{e11}
f = \left( \frac{eg}{\pi} \right)^2 ,
\ee
( $f^{-1}$ is the characteristic scale of length),
\be \label{e12}
\tilde{\cal B}(p) = \frac{1}{\sqrt{-p^2} + f} \; , \;
\tilde{\cal C}(p) = \frac{1}{\sqrt{-p^2}(\sqrt{-p^2}+f)} \; ,
\ee
\be \label{e13}
\tilde{\cal D}^{\mu\nu}(p) =
\frac{i \epsilon^{\mu\nu\lambda}p_{\lambda}}
{\sqrt{-p^2}(\sqrt{-p^2}+f)} \; .
\ee
To investigate the effective change of statistics, consider
a point source
\be \label{e14}
j^{\mu}(x) = ne \delta_0^{\mu} \delta^2 (\vec{x}) \; , \;
j_3^{\mu} (x) = n_3 \delta_0^{\mu} \delta^2 (\vec{x}) \; .
\ee
Substituting the Fourier transforms
$ \tilde{j}^{\mu} (p) = \frac{ne}{(2\pi)^3} \delta_0^{\mu}
\delta (p_0) , \tilde{j}_3^{\mu} (p) = \frac{n_3}{(2\pi)^3}
\delta_0^{\mu} \delta(p_0)$ in (\ref{e10}) and performing
the inverse transformation, we get
\be \label{e15}
 \begin{array}{rcl}
 A_0 (r) & = & ne \left[ \frac{1}{2\pi r} - \frac{f}{4}
 u(fr) \right] , \\ \\
 a_0 (r) & = & \frac{n_3 g^2}{4} u(fr) \end{array}
\ee
for the temporal components, and
\be \label{e16}
 \begin{array}{rcl}
 A_{\varphi} (r) & = & - \frac{sn_3}{2er} v(fr)  , \\ \\
 a_{\varphi} (r) & = & - \frac{sn}{2r} v(fr)  \end{array}
\ee
for the angular components ( $\varphi$ is the polar angle). Here
\be \label{e17}
u(x) = {\bf H}_0 (x) - Y_0 (x) = \frac{2}{\pi} x \, {}_1 F_2
\left( 1; \frac{3}{2} , \frac{3}{2} ; -\frac{x^2}{4} \right)
- Y_0 (x) ,
\ee
\be \label{e18}
v(x) = x \int_0^{\infty} \exp (-x \sinh t - t) \, dt \, .
\ee
The plots of $u(x)$ and $v(x)$ are displayed on Fig. 1.
The  temporal components correspond to the quasi-Coulomb
interaction  (note  that for $r \ll f^{-1}$ , $A_0$
behaves like $1/r$ and $a_0$ like $\ln r ,$  as  one should
expect in accordance with the form of the Lagrangian). The
effective change of statistics is due to the angular components.

For a general consideration, imagine a composite of $n_1$
fermions of the first sort ($\psi_1$) and $n_2$ of the
second sort ($\psi_2$). According to (\ref{e7}) and (\ref{e14}),
$n = n_1 + n_2$ and $n_3 = n_1 - n_2 .$  Interchanging  two
non-interacting such composites multiplies the wave function
by  the phase factor $\exp[i\pi n^2],$  so that the composites
themselves are fermions (bosons) for odd (even) $n .$
However, due to presence of the potentials (\ref{e16})
there appears an additional  phase  factor.  If the
composites are kept at a constant distance $r ,$ then
it  equals $\exp[i\pi\Delta(r)]$ where
\begin{eqnarray}
\Delta(r) & = & - \frac{1}{2\pi}
 \left[ ne A_{\varphi} (r) \cdot 2 \pi r +
 n_3 a_{\varphi} (r) \cdot 2 \pi r \right] \nonumber \\
 & = & snn_3 v(fr) \nonumber \\
 & = & s (n_1^2 - n_2^2) v(fr) . \label{e19}
\end{eqnarray}

At small distances  ($r \ll f^{-1}$) there is no  effective
change of statistics, since $v(0)=0 .$ At such distances
one should in general remember about the Coulomb
interaction; however, its energy, being of the order of
$e^2 /r + g^2 \ln r ,$ can always be made small enough by
the appropriate choice of parameters, while the function
$v(x)$ is parameterless. Therefore, at least from the
theoretical point of view, it is allowable not to take into
account this interaction \cite{r13}.

On the contrary, for $r \gg f^{-1}$ the  total phase change
is $\pi m ,$ where $m = (n_1 + n_2 )^2 +s(n_1^2 - n_2^2 )$
is  even  for any integer $n_1$  and $n_2 .$ Therefore
at large distances the considered composites always
behave as bosons.  At  intermediate  distances,
the behaviour is in some sense intermediate between
the two limiting cases; if $r$ is kept within a sufficiently
narrow  range  in  which $v(fr) \simeq \mbox{const} ,$
one effectively has anyons with the statistical  parameter
determined by (\ref{e19}). In the general case the situation
is more complicated. Let $n$ be odd so that the  composites
themselves are fermions. At high temperatures such that
$\lambda \ll \xi \;  (  \lambda$ is the thermal wavelength
and  $\xi \sim \rho^{-1/2}$ is the average interparticle
distance) a system of those behaves like Fermi gas for
$\lambda \ll f^{-1}$  and like Bose gas for
$\lambda \gg f^{-1}$  (although in both cases there is
only a small deviation of the equation of state from that
of the ideal gas) \cite{r13}. It seems plausible, however,
that it is by  the  relation  between $\lambda$  and $f^{-1}$
that the properties of the system are  determined for
$\lambda \gtrsim \xi$ as well. Therefore if the density is
fixed and the  temperature is being lowered, one should
expect the  behaviour  of  the system to change gradually
from Fermi-like to Bose-like. It would be an interesting
problem to study this process  in  more  details.  We
emphasize once more that since the model is P and  T-invariant,
the possibility of its realization in nature appears
considerably  more actual than for the usual P and
T-noninvariant Chern-Simons model.

E.V.G. would like to thank V.P.~Gusynin for a number of valuable
discussions. The work of S.V.M. was supported in part by a Soros
Foundation grant awarded by the American Physical Society.

\newpage
\begin{flushleft}
{\large Figure Caption} \\
\vskip 1.0cm
Fig.1. The functions $u(x)$ and $v(x)$. \\
\end{flushleft}
\newpage
\newcommand{\p}{\put}
\newcommand{\r}{\rule}
\newcommand{\m}{\makebox}
\newcommand{\y}{\usebox{\plotpoint}}
\setlength{\unitlength}{0.240900pt}
\ifx\plotpoint\undefined\newsavebox{\plotpoint}\fi
\sbox{\plotpoint}{\r[-0.175pt]{0.35pt}{0.35pt}}%
\begin{picture}(1500,900)(0,0)
\tenrm
\sbox{\plotpoint}{\r[-0.175pt]{0.35pt}{0.35pt}}%
\p(264,158){\r[-0.175pt]{282.335pt}{0.35pt}}
\p(264,158){\r[-0.175pt]{0.35pt}{151.526pt}}
\p(264,158){\r[-0.175pt]{4.818pt}{0.35pt}}
\p(242,158){\m(0,0)[r]{0}}
\p(1416,158){\r[-0.175pt]{4.818pt}{0.35pt}}
\p(264,284){\r[-0.175pt]{4.818pt}{0.35pt}}
\p(242,284){\m(0,0)[r]{0.2}}
\p(1416,284){\r[-0.175pt]{4.818pt}{0.35pt}}
\p(264,410){\r[-0.175pt]{4.818pt}{0.35pt}}
\p(242,410){\m(0,0)[r]{0.4}}
\p(1416,410){\r[-0.175pt]{4.818pt}{0.35pt}}
\p(264,535){\r[-0.175pt]{4.818pt}{0.35pt}}
\p(242,535){\m(0,0)[r]{0.6}}
\p(1416,535){\r[-0.175pt]{4.818pt}{0.35pt}}
\p(264,661){\r[-0.175pt]{4.818pt}{0.35pt}}
\p(242,661){\m(0,0)[r]{0.8}}
\p(1416,661){\r[-0.175pt]{4.818pt}{0.35pt}}
\p(264,787){\r[-0.175pt]{4.818pt}{0.35pt}}
\p(242,787){\m(0,0)[r]{1}}
\p(1416,787){\r[-0.175pt]{4.818pt}{0.35pt}}
\p(264,158){\r[-0.175pt]{0.35pt}{4.818pt}}
\p(264,113){\m(0,0){0}}
\p(264,767){\r[-0.175pt]{0.35pt}{4.818pt}}
\p(498,158){\r[-0.175pt]{0.35pt}{4.818pt}}
\p(498,113){\m(0,0){2}}
\p(498,767){\r[-0.175pt]{0.35pt}{4.818pt}}
\p(733,158){\r[-0.175pt]{0.35pt}{4.818pt}}
\p(733,113){\m(0,0){4}}
\p(733,767){\r[-0.175pt]{0.35pt}{4.818pt}}
\p(967,158){\r[-0.175pt]{0.35pt}{4.818pt}}
\p(967,113){\m(0,0){6}}
\p(967,767){\r[-0.175pt]{0.35pt}{4.818pt}}
\p(1202,158){\r[-0.175pt]{0.35pt}{4.818pt}}
\p(1202,113){\m(0,0){8}}
\p(1202,767){\r[-0.175pt]{0.35pt}{4.818pt}}
\p(1436,158){\r[-0.175pt]{0.35pt}{4.818pt}}
\p(1436,113){\m(0,0){10}}
\p(1436,767){\r[-0.175pt]{0.35pt}{4.818pt}}
\p(264,158){\r[-0.175pt]{282.335pt}{0.35pt}}
\p(1436,158){\r[-0.175pt]{0.35pt}{151.526pt}}
\p(264,787){\r[-0.175pt]{282.335pt}{0.35pt}}
\p(264,158){\r[-0.175pt]{0.35pt}{151.526pt}}
\p(1306,662){\m(0,0)[r]{u(x)}}
\p(1328,662){\r[-0.175pt]{15.899pt}{0.35pt}}
\p(299,787){\y}
\p(299,781){\r[-0.175pt]{0.35pt}{1.227pt}}
\p(300,776){\r[-0.175pt]{0.35pt}{1.227pt}}
\p(301,771){\r[-0.175pt]{0.35pt}{1.227pt}}
\p(302,766){\r[-0.175pt]{0.35pt}{1.227pt}}
\p(303,761){\r[-0.175pt]{0.35pt}{1.227pt}}
\p(304,756){\r[-0.175pt]{0.35pt}{1.227pt}}
\p(305,751){\r[-0.175pt]{0.35pt}{1.227pt}}
\p(306,746){\r[-0.175pt]{0.35pt}{1.227pt}}
\p(307,741){\r[-0.175pt]{0.35pt}{1.227pt}}
\p(308,736){\r[-0.175pt]{0.35pt}{1.227pt}}
\p(309,730){\r[-0.175pt]{0.35pt}{1.227pt}}
\p(310,725){\r[-0.175pt]{0.35pt}{1.227pt}}
\p(311,720){\r[-0.175pt]{0.35pt}{1.227pt}}
\p(312,715){\r[-0.175pt]{0.35pt}{1.227pt}}
\p(313,710){\r[-0.175pt]{0.35pt}{1.227pt}}
\p(314,705){\r[-0.175pt]{0.35pt}{1.227pt}}
\p(315,700){\r[-0.175pt]{0.35pt}{1.227pt}}
\p(316,695){\r[-0.175pt]{0.35pt}{1.227pt}}
\p(317,690){\r[-0.175pt]{0.35pt}{1.227pt}}
\p(318,685){\r[-0.175pt]{0.35pt}{1.227pt}}
\p(319,680){\r[-0.175pt]{0.35pt}{1.227pt}}
\p(320,675){\r[-0.175pt]{0.35pt}{0.988pt}}
\p(321,671){\r[-0.175pt]{0.35pt}{0.988pt}}
\p(322,667){\r[-0.175pt]{0.35pt}{0.988pt}}
\p(323,663){\r[-0.175pt]{0.35pt}{0.988pt}}
\p(324,659){\r[-0.175pt]{0.35pt}{0.988pt}}
\p(325,655){\r[-0.175pt]{0.35pt}{0.988pt}}
\p(326,651){\r[-0.175pt]{0.35pt}{0.988pt}}
\p(327,647){\r[-0.175pt]{0.35pt}{0.988pt}}
\p(328,643){\r[-0.175pt]{0.35pt}{0.988pt}}
\p(329,639){\r[-0.175pt]{0.35pt}{0.988pt}}
\p(330,634){\r[-0.175pt]{0.35pt}{0.988pt}}
\p(331,630){\r[-0.175pt]{0.35pt}{0.988pt}}
\p(332,626){\r[-0.175pt]{0.35pt}{0.988pt}}
\p(333,622){\r[-0.175pt]{0.35pt}{0.988pt}}
\p(334,618){\r[-0.175pt]{0.35pt}{0.988pt}}
\p(335,614){\r[-0.175pt]{0.35pt}{0.988pt}}
\p(336,610){\r[-0.175pt]{0.35pt}{0.988pt}}
\p(337,606){\r[-0.175pt]{0.35pt}{0.988pt}}
\p(338,602){\r[-0.175pt]{0.35pt}{0.988pt}}
\p(339,598){\r[-0.175pt]{0.35pt}{0.988pt}}
\p(340,595){\r[-0.175pt]{0.35pt}{0.723pt}}
\p(341,592){\r[-0.175pt]{0.35pt}{0.723pt}}
\p(342,589){\r[-0.175pt]{0.35pt}{0.723pt}}
\p(343,586){\r[-0.175pt]{0.35pt}{0.723pt}}
\p(344,583){\r[-0.175pt]{0.35pt}{0.723pt}}
\p(345,580){\r[-0.175pt]{0.35pt}{0.723pt}}
\p(346,577){\r[-0.175pt]{0.35pt}{0.723pt}}
\p(347,574){\r[-0.175pt]{0.35pt}{0.723pt}}
\p(348,571){\r[-0.175pt]{0.35pt}{0.723pt}}
\p(349,568){\r[-0.175pt]{0.35pt}{0.723pt}}
\p(350,565){\r[-0.175pt]{0.35pt}{0.723pt}}
\p(351,562){\r[-0.175pt]{0.35pt}{0.723pt}}
\p(352,559){\r[-0.175pt]{0.35pt}{0.723pt}}
\p(353,556){\r[-0.175pt]{0.35pt}{0.723pt}}
\p(354,553){\r[-0.175pt]{0.35pt}{0.723pt}}
\p(355,550){\r[-0.175pt]{0.35pt}{0.723pt}}
\p(356,547){\r[-0.175pt]{0.35pt}{0.723pt}}
\p(357,544){\r[-0.175pt]{0.35pt}{0.723pt}}
\p(358,541){\r[-0.175pt]{0.35pt}{0.723pt}}
\p(359,538){\r[-0.175pt]{0.35pt}{0.723pt}}
\p(360,535){\r[-0.175pt]{0.35pt}{0.723pt}}
\p(361,532){\r[-0.175pt]{0.35pt}{0.602pt}}
\p(362,530){\r[-0.175pt]{0.35pt}{0.602pt}}
\p(363,527){\r[-0.175pt]{0.35pt}{0.602pt}}
\p(364,525){\r[-0.175pt]{0.35pt}{0.602pt}}
\p(365,522){\r[-0.175pt]{0.35pt}{0.602pt}}
\p(366,520){\r[-0.175pt]{0.35pt}{0.602pt}}
\p(367,517){\r[-0.175pt]{0.35pt}{0.602pt}}
\p(368,515){\r[-0.175pt]{0.35pt}{0.602pt}}
\p(369,512){\r[-0.175pt]{0.35pt}{0.602pt}}
\p(370,510){\r[-0.175pt]{0.35pt}{0.602pt}}
\p(371,507){\r[-0.175pt]{0.35pt}{0.602pt}}
\p(372,505){\r[-0.175pt]{0.35pt}{0.602pt}}
\p(373,502){\r[-0.175pt]{0.35pt}{0.602pt}}
\p(374,500){\r[-0.175pt]{0.35pt}{0.602pt}}
\p(375,497){\r[-0.175pt]{0.35pt}{0.602pt}}
\p(376,495){\r[-0.175pt]{0.35pt}{0.602pt}}
\p(377,492){\r[-0.175pt]{0.35pt}{0.602pt}}
\p(378,490){\r[-0.175pt]{0.35pt}{0.602pt}}
\p(379,487){\r[-0.175pt]{0.35pt}{0.602pt}}
\p(380,485){\r[-0.175pt]{0.35pt}{0.602pt}}
\p(381,483){\r[-0.175pt]{0.35pt}{0.377pt}}
\p(382,481){\r[-0.175pt]{0.35pt}{0.377pt}}
\p(383,480){\r[-0.175pt]{0.35pt}{0.377pt}}
\p(384,478){\r[-0.175pt]{0.35pt}{0.377pt}}
\p(385,477){\r[-0.175pt]{0.35pt}{0.377pt}}
\p(386,475){\r[-0.175pt]{0.35pt}{0.377pt}}
\p(387,474){\r[-0.175pt]{0.35pt}{0.377pt}}
\p(388,472){\r[-0.175pt]{0.35pt}{0.377pt}}
\p(389,470){\r[-0.175pt]{0.35pt}{0.377pt}}
\p(390,469){\r[-0.175pt]{0.35pt}{0.377pt}}
\p(391,467){\r[-0.175pt]{0.35pt}{0.377pt}}
\p(392,466){\r[-0.175pt]{0.35pt}{0.377pt}}
\p(393,464){\r[-0.175pt]{0.35pt}{0.377pt}}
\p(394,463){\r[-0.175pt]{0.35pt}{0.377pt}}
\p(395,461){\r[-0.175pt]{0.35pt}{0.377pt}}
\p(396,459){\r[-0.175pt]{0.35pt}{0.377pt}}
\p(397,458){\r[-0.175pt]{0.35pt}{0.377pt}}
\p(398,456){\r[-0.175pt]{0.35pt}{0.377pt}}
\p(399,455){\r[-0.175pt]{0.35pt}{0.377pt}}
\p(400,453){\r[-0.175pt]{0.35pt}{0.377pt}}
\p(401,452){\r[-0.175pt]{0.35pt}{0.377pt}}
\p(402,450){\r[-0.175pt]{0.35pt}{0.377pt}}
\p(403,448){\r[-0.175pt]{0.35pt}{0.377pt}}
\p(404,447){\r[-0.175pt]{0.35pt}{0.377pt}}
\p(405,445){\r[-0.175pt]{0.35pt}{0.377pt}}
\p(406,444){\r[-0.175pt]{0.35pt}{0.377pt}}
\p(407,442){\r[-0.175pt]{0.35pt}{0.377pt}}
\p(408,441){\r[-0.175pt]{0.35pt}{0.377pt}}
\p(409,439){\r[-0.175pt]{0.35pt}{0.377pt}}
\p(410,438){\r[-0.175pt]{0.35pt}{0.377pt}}
\p(411,436){\r[-0.175pt]{0.35pt}{0.374pt}}
\p(412,434){\r[-0.175pt]{0.35pt}{0.374pt}}
\p(413,433){\r[-0.175pt]{0.35pt}{0.374pt}}
\p(414,431){\r[-0.175pt]{0.35pt}{0.374pt}}
\p(415,430){\r[-0.175pt]{0.35pt}{0.374pt}}
\p(416,428){\r[-0.175pt]{0.35pt}{0.374pt}}
\p(417,427){\r[-0.175pt]{0.35pt}{0.374pt}}
\p(418,425){\r[-0.175pt]{0.35pt}{0.374pt}}
\p(419,424){\r[-0.175pt]{0.35pt}{0.374pt}}
\p(420,422){\r[-0.175pt]{0.35pt}{0.374pt}}
\p(421,420){\r[-0.175pt]{0.35pt}{0.374pt}}
\p(422,419){\r[-0.175pt]{0.35pt}{0.374pt}}
\p(423,417){\r[-0.175pt]{0.35pt}{0.374pt}}
\p(424,416){\r[-0.175pt]{0.35pt}{0.374pt}}
\p(425,414){\r[-0.175pt]{0.35pt}{0.374pt}}
\p(426,413){\r[-0.175pt]{0.35pt}{0.374pt}}
\p(427,411){\r[-0.175pt]{0.35pt}{0.374pt}}
\p(428,410){\r[-0.175pt]{0.35pt}{0.374pt}}
\p(429,408){\r[-0.175pt]{0.35pt}{0.374pt}}
\p(430,406){\r[-0.175pt]{0.35pt}{0.374pt}}
\p(431,405){\r[-0.175pt]{0.35pt}{0.374pt}}
\p(432,403){\r[-0.175pt]{0.35pt}{0.374pt}}
\p(433,402){\r[-0.175pt]{0.35pt}{0.374pt}}
\p(434,400){\r[-0.175pt]{0.35pt}{0.374pt}}
\p(435,399){\r[-0.175pt]{0.35pt}{0.374pt}}
\p(436,397){\r[-0.175pt]{0.35pt}{0.374pt}}
\p(437,396){\r[-0.175pt]{0.35pt}{0.374pt}}
\p(438,394){\r[-0.175pt]{0.35pt}{0.374pt}}
\p(439,393){\r[-0.175pt]{0.35pt}{0.374pt}}
\p(440,393){\y}
\p(441,392){\y}
\p(442,391){\y}
\p(443,390){\y}
\p(444,389){\y}
\p(445,388){\y}
\p(446,387){\y}
\p(447,386){\y}
\p(448,385){\y}
\p(450,384){\y}
\p(451,383){\y}
\p(452,382){\y}
\p(453,381){\y}
\p(454,380){\y}
\p(455,379){\y}
\p(456,378){\y}
\p(457,377){\y}
\p(458,376){\y}
\p(460,375){\y}
\p(461,374){\y}
\p(462,373){\y}
\p(463,372){\y}
\p(464,371){\y}
\p(465,370){\y}
\p(466,369){\y}
\p(467,368){\y}
\p(469,367){\r[-0.175pt]{0.388pt}{0.35pt}}
\p(470,366){\r[-0.175pt]{0.388pt}{0.35pt}}
\p(472,365){\r[-0.175pt]{0.388pt}{0.35pt}}
\p(473,364){\r[-0.175pt]{0.388pt}{0.35pt}}
\p(475,363){\r[-0.175pt]{0.388pt}{0.35pt}}
\p(477,362){\r[-0.175pt]{0.388pt}{0.35pt}}
\p(478,361){\r[-0.175pt]{0.388pt}{0.35pt}}
\p(480,360){\r[-0.175pt]{0.388pt}{0.35pt}}
\p(481,359){\r[-0.175pt]{0.388pt}{0.35pt}}
\p(483,358){\r[-0.175pt]{0.388pt}{0.35pt}}
\p(485,357){\r[-0.175pt]{0.388pt}{0.35pt}}
\p(486,356){\r[-0.175pt]{0.388pt}{0.35pt}}
\p(488,355){\r[-0.175pt]{0.388pt}{0.35pt}}
\p(489,354){\r[-0.175pt]{0.388pt}{0.35pt}}
\p(491,353){\r[-0.175pt]{0.388pt}{0.35pt}}
\p(493,352){\r[-0.175pt]{0.388pt}{0.35pt}}
\p(494,351){\r[-0.175pt]{0.388pt}{0.35pt}}
\p(496,350){\r[-0.175pt]{0.388pt}{0.35pt}}
\p(498,349){\r[-0.175pt]{0.418pt}{0.35pt}}
\p(499,348){\r[-0.175pt]{0.418pt}{0.35pt}}
\p(501,347){\r[-0.175pt]{0.418pt}{0.35pt}}
\p(503,346){\r[-0.175pt]{0.418pt}{0.35pt}}
\p(504,345){\r[-0.175pt]{0.418pt}{0.35pt}}
\p(506,344){\r[-0.175pt]{0.418pt}{0.35pt}}
\p(508,343){\r[-0.175pt]{0.418pt}{0.35pt}}
\p(510,342){\r[-0.175pt]{0.418pt}{0.35pt}}
\p(511,341){\r[-0.175pt]{0.418pt}{0.35pt}}
\p(513,340){\r[-0.175pt]{0.418pt}{0.35pt}}
\p(515,339){\r[-0.175pt]{0.418pt}{0.35pt}}
\p(517,338){\r[-0.175pt]{0.418pt}{0.35pt}}
\p(518,337){\r[-0.175pt]{0.418pt}{0.35pt}}
\p(520,336){\r[-0.175pt]{0.418pt}{0.35pt}}
\p(522,335){\r[-0.175pt]{0.418pt}{0.35pt}}
\p(524,334){\r[-0.175pt]{0.418pt}{0.35pt}}
\p(525,333){\r[-0.175pt]{0.418pt}{0.35pt}}
\p(527,332){\r[-0.175pt]{0.418pt}{0.35pt}}
\p(529,331){\r[-0.175pt]{0.418pt}{0.35pt}}
\p(530,330){\r[-0.175pt]{0.418pt}{0.35pt}}
\p(532,329){\r[-0.175pt]{0.418pt}{0.35pt}}
\p(534,328){\r[-0.175pt]{0.418pt}{0.35pt}}
\p(536,327){\r[-0.175pt]{0.418pt}{0.35pt}}
\p(537,326){\r[-0.175pt]{0.418pt}{0.35pt}}
\p(539,325){\r[-0.175pt]{0.418pt}{0.35pt}}
\p(541,324){\r[-0.175pt]{0.418pt}{0.35pt}}
\p(543,323){\r[-0.175pt]{0.418pt}{0.35pt}}
\p(544,322){\r[-0.175pt]{0.418pt}{0.35pt}}
\p(546,321){\r[-0.175pt]{0.418pt}{0.35pt}}
\p(548,320){\r[-0.175pt]{0.418pt}{0.35pt}}
\p(550,319){\r[-0.175pt]{0.418pt}{0.35pt}}
\p(551,318){\r[-0.175pt]{0.418pt}{0.35pt}}
\p(553,317){\r[-0.175pt]{0.418pt}{0.35pt}}
\p(555,316){\r[-0.175pt]{0.418pt}{0.35pt}}
\p(556,315){\r[-0.175pt]{0.646pt}{0.35pt}}
\p(559,314){\r[-0.175pt]{0.646pt}{0.35pt}}
\p(562,313){\r[-0.175pt]{0.646pt}{0.35pt}}
\p(565,312){\r[-0.175pt]{0.646pt}{0.35pt}}
\p(567,311){\r[-0.175pt]{0.646pt}{0.35pt}}
\p(570,310){\r[-0.175pt]{0.646pt}{0.35pt}}
\p(573,309){\r[-0.175pt]{0.646pt}{0.35pt}}
\p(575,308){\r[-0.175pt]{0.646pt}{0.35pt}}
\p(578,307){\r[-0.175pt]{0.646pt}{0.35pt}}
\p(581,306){\r[-0.175pt]{0.646pt}{0.35pt}}
\p(583,305){\r[-0.175pt]{0.646pt}{0.35pt}}
\p(586,304){\r[-0.175pt]{0.646pt}{0.35pt}}
\p(589,303){\r[-0.175pt]{0.646pt}{0.35pt}}
\p(591,302){\r[-0.175pt]{0.646pt}{0.35pt}}
\p(594,301){\r[-0.175pt]{0.646pt}{0.35pt}}
\p(597,300){\r[-0.175pt]{0.646pt}{0.35pt}}
\p(599,299){\r[-0.175pt]{0.646pt}{0.35pt}}
\p(602,298){\r[-0.175pt]{0.646pt}{0.35pt}}
\p(605,297){\r[-0.175pt]{0.646pt}{0.35pt}}
\p(607,296){\r[-0.175pt]{0.646pt}{0.35pt}}
\p(610,295){\r[-0.175pt]{0.646pt}{0.35pt}}
\p(613,294){\r[-0.175pt]{0.646pt}{0.35pt}}
\p(616,293){\r[-0.175pt]{1.007pt}{0.35pt}}
\p(620,292){\r[-0.175pt]{1.007pt}{0.35pt}}
\p(624,291){\r[-0.175pt]{1.007pt}{0.35pt}}
\p(628,290){\r[-0.175pt]{1.007pt}{0.35pt}}
\p(632,289){\r[-0.175pt]{1.007pt}{0.35pt}}
\p(636,288){\r[-0.175pt]{1.007pt}{0.35pt}}
\p(641,287){\r[-0.175pt]{1.007pt}{0.35pt}}
\p(645,286){\r[-0.175pt]{1.007pt}{0.35pt}}
\p(649,285){\r[-0.175pt]{1.007pt}{0.35pt}}
\p(653,284){\r[-0.175pt]{1.007pt}{0.35pt}}
\p(657,283){\r[-0.175pt]{1.007pt}{0.35pt}}
\p(661,282){\r[-0.175pt]{1.007pt}{0.35pt}}
\p(666,281){\r[-0.175pt]{1.007pt}{0.35pt}}
\p(670,280){\r[-0.175pt]{1.007pt}{0.35pt}}
\p(674,279){\r[-0.175pt]{1.007pt}{0.35pt}}
\p(678,278){\r[-0.175pt]{1.007pt}{0.35pt}}
\p(682,277){\r[-0.175pt]{1.007pt}{0.35pt}}
\p(687,276){\r[-0.175pt]{1.007pt}{0.35pt}}
\p(691,275){\r[-0.175pt]{1.007pt}{0.35pt}}
\p(695,274){\r[-0.175pt]{1.007pt}{0.35pt}}
\p(699,273){\r[-0.175pt]{1.007pt}{0.35pt}}
\p(703,272){\r[-0.175pt]{1.007pt}{0.35pt}}
\p(707,271){\r[-0.175pt]{1.007pt}{0.35pt}}
\p(712,270){\r[-0.175pt]{1.007pt}{0.35pt}}
\p(716,269){\r[-0.175pt]{1.007pt}{0.35pt}}
\p(720,268){\r[-0.175pt]{1.007pt}{0.35pt}}
\p(724,267){\r[-0.175pt]{1.007pt}{0.35pt}}
\p(728,266){\r[-0.175pt]{1.007pt}{0.35pt}}
\p(733,265){\r[-0.175pt]{1.483pt}{0.35pt}}
\p(739,264){\r[-0.175pt]{1.483pt}{0.35pt}}
\p(745,263){\r[-0.175pt]{1.483pt}{0.35pt}}
\p(751,262){\r[-0.175pt]{1.483pt}{0.35pt}}
\p(757,261){\r[-0.175pt]{1.483pt}{0.35pt}}
\p(763,260){\r[-0.175pt]{1.483pt}{0.35pt}}
\p(769,259){\r[-0.175pt]{1.483pt}{0.35pt}}
\p(776,258){\r[-0.175pt]{1.483pt}{0.35pt}}
\p(782,257){\r[-0.175pt]{1.483pt}{0.35pt}}
\p(788,256){\r[-0.175pt]{1.483pt}{0.35pt}}
\p(794,255){\r[-0.175pt]{1.483pt}{0.35pt}}
\p(800,254){\r[-0.175pt]{1.483pt}{0.35pt}}
\p(806,253){\r[-0.175pt]{1.483pt}{0.35pt}}
\p(813,252){\r[-0.175pt]{1.483pt}{0.35pt}}
\p(819,251){\r[-0.175pt]{1.483pt}{0.35pt}}
\p(825,250){\r[-0.175pt]{1.483pt}{0.35pt}}
\p(831,249){\r[-0.175pt]{1.483pt}{0.35pt}}
\p(837,248){\r[-0.175pt]{1.483pt}{0.35pt}}
\p(843,247){\r[-0.175pt]{1.483pt}{0.35pt}}
\p(850,246){\r[-0.175pt]{2.168pt}{0.35pt}}
\p(859,245){\r[-0.175pt]{2.168pt}{0.35pt}}
\p(868,244){\r[-0.175pt]{2.168pt}{0.35pt}}
\p(877,243){\r[-0.175pt]{2.168pt}{0.35pt}}
\p(886,242){\r[-0.175pt]{2.168pt}{0.35pt}}
\p(895,241){\r[-0.175pt]{2.168pt}{0.35pt}}
\p(904,240){\r[-0.175pt]{2.168pt}{0.35pt}}
\p(913,239){\r[-0.175pt]{2.168pt}{0.35pt}}
\p(922,238){\r[-0.175pt]{2.168pt}{0.35pt}}
\p(931,237){\r[-0.175pt]{2.168pt}{0.35pt}}
\p(940,236){\r[-0.175pt]{2.168pt}{0.35pt}}
\p(949,235){\r[-0.175pt]{2.168pt}{0.35pt}}
\p(958,234){\r[-0.175pt]{2.168pt}{0.35pt}}
\p(967,233){\r[-0.175pt]{4.698pt}{0.35pt}}
\p(986,232){\r[-0.175pt]{4.698pt}{0.35pt}}
\p(1006,231){\r[-0.175pt]{4.698pt}{0.35pt}}
\p(1025,230){\r[-0.175pt]{4.698pt}{0.35pt}}
\p(1045,229){\r[-0.175pt]{4.698pt}{0.35pt}}
\p(1064,228){\r[-0.175pt]{4.698pt}{0.35pt}}
\p(1084,227){\r[-0.175pt]{9.475pt}{0.35pt}}
\p(1123,226){\r[-0.175pt]{9.475pt}{0.35pt}}
\p(1162,225){\r[-0.175pt]{9.475pt}{0.35pt}}
\p(1202,224){\r[-0.175pt]{9.395pt}{0.35pt}}
\p(1241,223){\r[-0.175pt]{9.395pt}{0.35pt}}
\p(1280,222){\r[-0.175pt]{9.395pt}{0.35pt}}
\p(1319,221){\r[-0.175pt]{14.093pt}{0.35pt}}
\p(1377,220){\r[-0.175pt]{14.093pt}{0.35pt}}
\sbox{\plotpoint}{\r[-0.35pt]{0.7pt}{0.7pt}}%
\p(1306,617){\m(0,0)[r]{v(x)}}
\p(1328,617){\r[-0.35pt]{15.899pt}{0.7pt}}
\p(264,158){\y}
\p(264,158){\r[-0.35pt]{0.7pt}{1.084pt}}
\p(265,162){\r[-0.35pt]{0.7pt}{1.084pt}}
\p(266,167){\r[-0.35pt]{0.7pt}{1.084pt}}
\p(267,171){\r[-0.35pt]{0.7pt}{1.084pt}}
\p(268,176){\r[-0.35pt]{0.7pt}{1.084pt}}
\p(269,180){\r[-0.35pt]{0.7pt}{1.084pt}}
\p(270,185){\r[-0.35pt]{0.7pt}{1.084pt}}
\p(271,189){\r[-0.35pt]{0.7pt}{1.084pt}}
\p(272,194){\r[-0.35pt]{0.7pt}{1.084pt}}
\p(273,198){\r[-0.35pt]{0.7pt}{1.084pt}}
\p(274,203){\r[-0.35pt]{0.7pt}{1.084pt}}
\p(275,207){\r[-0.35pt]{0.7pt}{1.084pt}}
\p(276,212){\r[-0.35pt]{0.7pt}{0.920pt}}
\p(277,215){\r[-0.35pt]{0.7pt}{0.920pt}}
\p(278,219){\r[-0.35pt]{0.7pt}{0.920pt}}
\p(279,223){\r[-0.35pt]{0.7pt}{0.920pt}}
\p(280,227){\r[-0.35pt]{0.7pt}{0.920pt}}
\p(281,231){\r[-0.35pt]{0.7pt}{0.920pt}}
\p(282,234){\r[-0.35pt]{0.7pt}{0.920pt}}
\p(283,238){\r[-0.35pt]{0.7pt}{0.920pt}}
\p(284,242){\r[-0.35pt]{0.7pt}{0.920pt}}
\p(285,246){\r[-0.35pt]{0.7pt}{0.920pt}}
\p(286,250){\r[-0.35pt]{0.7pt}{0.920pt}}
\p(287,253){\r[-0.35pt]{0.7pt}{0.723pt}}
\p(288,257){\r[-0.35pt]{0.7pt}{0.723pt}}
\p(289,260){\r[-0.35pt]{0.7pt}{0.723pt}}
\p(290,263){\r[-0.35pt]{0.7pt}{0.723pt}}
\p(291,266){\r[-0.35pt]{0.7pt}{0.723pt}}
\p(292,269){\r[-0.35pt]{0.7pt}{0.723pt}}
\p(293,272){\r[-0.35pt]{0.7pt}{0.723pt}}
\p(294,275){\r[-0.35pt]{0.7pt}{0.723pt}}
\p(295,278){\r[-0.35pt]{0.7pt}{0.723pt}}
\p(296,281){\r[-0.35pt]{0.7pt}{0.723pt}}
\p(297,284){\r[-0.35pt]{0.7pt}{0.723pt}}
\p(298,287){\r[-0.35pt]{0.7pt}{0.723pt}}
\p(299,290){\y}
\p(300,292){\y}
\p(301,295){\y}
\p(302,297){\y}
\p(303,300){\y}
\p(304,302){\y}
\p(305,305){\y}
\p(306,308){\y}
\p(307,310){\y}
\p(308,313){\y}
\p(309,315){\y}
\p(310,318){\y}
\p(311,321){\y}
\p(312,323){\y}
\p(313,325){\y}
\p(314,328){\y}
\p(315,330){\y}
\p(316,332){\y}
\p(317,335){\y}
\p(318,337){\y}
\p(319,339){\y}
\p(320,342){\y}
\p(321,344){\y}
\p(322,346){\y}
\p(323,349){\y}
\p(324,351){\y}
\p(325,353){\y}
\p(326,355){\y}
\p(327,357){\y}
\p(328,359){\y}
\p(329,362){\y}
\p(330,364){\y}
\p(331,366){\y}
\p(332,368){\y}
\p(333,370){\y}
\p(334,373){\y}
\p(335,374){\y}
\p(336,376){\y}
\p(337,378){\y}
\p(338,380){\y}
\p(339,382){\y}
\p(340,384){\y}
\p(341,385){\y}
\p(342,387){\y}
\p(343,389){\y}
\p(344,391){\y}
\p(345,393){\y}
\p(346,395){\y}
\p(347,396){\y}
\p(348,398){\y}
\p(349,399){\y}
\p(350,401){\y}
\p(351,402){\y}
\p(352,404){\y}
\p(353,406){\y}
\p(354,407){\y}
\p(355,409){\y}
\p(356,410){\y}
\p(357,412){\y}
\p(358,414){\y}
\p(359,415){\y}
\p(360,417){\y}
\p(361,418){\y}
\p(362,420){\y}
\p(363,422){\y}
\p(364,423){\y}
\p(365,425){\y}
\p(366,427){\y}
\p(367,428){\y}
\p(368,430){\y}
\p(369,431){\y}
\p(370,433){\y}
\p(371,434){\y}
\p(372,436){\y}
\p(373,437){\y}
\p(374,438){\y}
\p(375,440){\y}
\p(376,441){\y}
\p(377,442){\y}
\p(378,444){\y}
\p(379,445){\y}
\p(380,446){\y}
\p(381,448){\y}
\p(382,449){\y}
\p(383,450){\y}
\p(384,451){\y}
\p(385,453){\y}
\p(386,454){\y}
\p(387,455){\y}
\p(388,456){\y}
\p(389,458){\y}
\p(390,459){\y}
\p(391,460){\y}
\p(392,461){\y}
\p(393,463){\y}
\p(394,464){\y}
\p(395,465){\y}
\p(396,466){\y}
\p(397,467){\y}
\p(398,468){\y}
\p(399,469){\y}
\p(400,470){\y}
\p(401,471){\y}
\p(402,472){\y}
\p(403,473){\y}
\p(404,474){\y}
\p(405,476){\y}
\p(406,477){\y}
\p(407,478){\y}
\p(408,479){\y}
\p(409,480){\y}
\p(410,481){\y}
\p(411,483){\y}
\p(412,484){\y}
\p(413,485){\y}
\p(414,486){\y}
\p(415,487){\y}
\p(416,489){\y}
\p(417,490){\y}
\p(418,491){\y}
\p(419,492){\y}
\p(420,493){\y}
\p(421,494){\y}
\p(422,495){\y}
\p(423,496){\y}
\p(424,497){\y}
\p(425,498){\y}
\p(426,499){\y}
\p(427,500){\y}
\p(428,501){\y}
\p(429,502){\y}
\p(430,503){\y}
\p(431,504){\y}
\p(432,505){\y}
\p(434,506){\y}
\p(435,507){\y}
\p(436,508){\y}
\p(437,509){\y}
\p(438,510){\y}
\p(440,511){\y}
\p(441,512){\y}
\p(442,513){\y}
\p(443,514){\y}
\p(444,515){\y}
\p(446,516){\y}
\p(447,517){\y}
\p(448,518){\y}
\p(449,519){\y}
\p(450,520){\y}
\p(452,521){\y}
\p(453,522){\y}
\p(454,523){\y}
\p(455,524){\y}
\p(456,525){\y}
\p(457,526){\y}
\p(458,527){\y}
\p(459,528){\y}
\p(460,529){\y}
\p(461,530){\y}
\p(463,531){\y}
\p(464,532){\y}
\p(466,533){\y}
\p(467,534){\y}
\p(469,535){\y}
\p(470,536){\y}
\p(472,537){\y}
\p(473,538){\y}
\p(475,539){\y}
\p(476,540){\y}
\p(477,541){\y}
\p(479,542){\y}
\p(480,543){\y}
\p(481,544){\y}
\p(483,545){\y}
\p(484,546){\y}
\p(485,547){\y}
\p(487,548){\y}
\p(488,549){\y}
\p(490,550){\y}
\p(491,551){\y}
\p(493,552){\y}
\p(494,553){\y}
\p(496,554){\y}
\p(498,555){\y}
\p(499,556){\y}
\p(501,557){\y}
\p(502,558){\y}
\p(504,559){\y}
\p(505,560){\y}
\p(507,561){\y}
\p(508,562){\y}
\p(510,563){\y}
\p(512,564){\y}
\p(514,565){\y}
\p(516,566){\y}
\p(518,567){\y}
\p(520,568){\y}
\p(522,569){\y}
\p(523,570){\y}
\p(525,571){\y}
\p(527,572){\y}
\p(528,573){\y}
\p(530,574){\y}
\p(532,575){\y}
\p(534,576){\y}
\p(535,577){\y}
\p(537,578){\y}
\p(539,579){\y}
\p(541,580){\y}
\p(543,581){\y}
\p(544,582){\y}
\p(547,583){\y}
\p(549,584){\y}
\p(551,585){\y}
\p(553,586){\y}
\p(555,587){\y}
\p(557,588){\y}
\p(559,589){\y}
\p(561,590){\y}
\p(564,591){\y}
\p(566,592){\y}
\p(569,593){\y}
\p(571,594){\y}
\p(573,595){\y}
\p(575,596){\y}
\p(577,597){\y}
\p(580,598){\y}
\p(582,599){\y}
\p(584,600){\y}
\p(587,601){\y}
\p(589,602){\y}
\p(592,603){\y}
\p(594,604){\y}
\p(596,605){\y}
\p(599,606){\y}
\p(601,607){\y}
\p(604,608){\r[-0.35pt]{0.723pt}{0.7pt}}
\p(607,609){\r[-0.35pt]{0.723pt}{0.7pt}}
\p(610,610){\r[-0.35pt]{0.723pt}{0.7pt}}
\p(613,611){\r[-0.35pt]{0.723pt}{0.7pt}}
\p(616,612){\y}
\p(618,613){\y}
\p(620,614){\y}
\p(622,615){\y}
\p(624,616){\y}
\p(627,617){\r[-0.35pt]{0.723pt}{0.7pt}}
\p(630,618){\r[-0.35pt]{0.723pt}{0.7pt}}
\p(633,619){\r[-0.35pt]{0.723pt}{0.7pt}}
\p(636,620){\r[-0.35pt]{0.723pt}{0.7pt}}
\p(639,621){\r[-0.35pt]{0.723pt}{0.7pt}}
\p(642,622){\r[-0.35pt]{0.723pt}{0.7pt}}
\p(645,623){\r[-0.35pt]{0.723pt}{0.7pt}}
\p(648,624){\r[-0.35pt]{0.723pt}{0.7pt}}
\p(651,625){\r[-0.35pt]{0.883pt}{0.7pt}}
\p(654,626){\r[-0.35pt]{0.883pt}{0.7pt}}
\p(658,627){\r[-0.35pt]{0.883pt}{0.7pt}}
\p(662,628){\r[-0.35pt]{0.723pt}{0.7pt}}
\p(665,629){\r[-0.35pt]{0.723pt}{0.7pt}}
\p(668,630){\r[-0.35pt]{0.723pt}{0.7pt}}
\p(671,631){\r[-0.35pt]{0.723pt}{0.7pt}}
\p(674,632){\r[-0.35pt]{0.964pt}{0.7pt}}
\p(678,633){\r[-0.35pt]{0.964pt}{0.7pt}}
\p(682,634){\r[-0.35pt]{0.964pt}{0.7pt}}
\p(686,635){\r[-0.35pt]{0.723pt}{0.7pt}}
\p(689,636){\r[-0.35pt]{0.723pt}{0.7pt}}
\p(692,637){\r[-0.35pt]{0.723pt}{0.7pt}}
\p(695,638){\r[-0.35pt]{0.723pt}{0.7pt}}
\p(698,639){\r[-0.35pt]{0.883pt}{0.7pt}}
\p(701,640){\r[-0.35pt]{0.883pt}{0.7pt}}
\p(705,641){\r[-0.35pt]{0.883pt}{0.7pt}}
\p(709,642){\r[-0.35pt]{0.964pt}{0.7pt}}
\p(713,643){\r[-0.35pt]{0.964pt}{0.7pt}}
\p(717,644){\r[-0.35pt]{0.964pt}{0.7pt}}
\p(721,645){\r[-0.35pt]{0.964pt}{0.7pt}}
\p(725,646){\r[-0.35pt]{0.964pt}{0.7pt}}
\p(729,647){\r[-0.35pt]{0.964pt}{0.7pt}}
\p(733,648){\r[-0.35pt]{1.132pt}{0.7pt}}
\p(737,649){\r[-0.35pt]{1.132pt}{0.7pt}}
\p(742,650){\r[-0.35pt]{1.132pt}{0.7pt}}
\p(747,651){\r[-0.35pt]{1.132pt}{0.7pt}}
\p(751,652){\r[-0.35pt]{1.132pt}{0.7pt}}
\p(756,653){\r[-0.35pt]{1.132pt}{0.7pt}}
\p(761,654){\r[-0.35pt]{1.132pt}{0.7pt}}
\p(765,655){\r[-0.35pt]{1.132pt}{0.7pt}}
\p(770,656){\r[-0.35pt]{1.132pt}{0.7pt}}
\p(775,657){\r[-0.35pt]{1.132pt}{0.7pt}}
\p(780,658){\r[-0.35pt]{1.258pt}{0.7pt}}
\p(785,659){\r[-0.35pt]{1.258pt}{0.7pt}}
\p(790,660){\r[-0.35pt]{1.258pt}{0.7pt}}
\p(795,661){\r[-0.35pt]{1.258pt}{0.7pt}}
\p(800,662){\r[-0.35pt]{1.258pt}{0.7pt}}
\p(806,663){\r[-0.35pt]{1.258pt}{0.7pt}}
\p(811,664){\r[-0.35pt]{1.258pt}{0.7pt}}
\p(816,665){\r[-0.35pt]{1.258pt}{0.7pt}}
\p(821,666){\r[-0.35pt]{1.258pt}{0.7pt}}
\p(827,667){\r[-0.35pt]{1.385pt}{0.7pt}}
\p(832,668){\r[-0.35pt]{1.385pt}{0.7pt}}
\p(838,669){\r[-0.35pt]{1.385pt}{0.7pt}}
\p(844,670){\r[-0.35pt]{1.385pt}{0.7pt}}
\p(850,671){\r[-0.35pt]{1.385pt}{0.7pt}}
\p(855,672){\r[-0.35pt]{1.385pt}{0.7pt}}
\p(861,673){\r[-0.35pt]{1.385pt}{0.7pt}}
\p(867,674){\r[-0.35pt]{1.385pt}{0.7pt}}
\p(873,675){\r[-0.35pt]{1.617pt}{0.7pt}}
\p(879,676){\r[-0.35pt]{1.617pt}{0.7pt}}
\p(886,677){\r[-0.35pt]{1.617pt}{0.7pt}}
\p(893,678){\r[-0.35pt]{1.617pt}{0.7pt}}
\p(899,679){\r[-0.35pt]{1.617pt}{0.7pt}}
\p(906,680){\r[-0.35pt]{1.617pt}{0.7pt}}
\p(913,681){\r[-0.35pt]{1.617pt}{0.7pt}}
\p(920,682){\r[-0.35pt]{1.887pt}{0.7pt}}
\p(927,683){\r[-0.35pt]{1.887pt}{0.7pt}}
\p(935,684){\r[-0.35pt]{1.887pt}{0.7pt}}
\p(943,685){\r[-0.35pt]{1.887pt}{0.7pt}}
\p(951,686){\r[-0.35pt]{1.887pt}{0.7pt}}
\p(959,687){\r[-0.35pt]{1.887pt}{0.7pt}}
\p(966,688){\r[-0.35pt]{1.887pt}{0.7pt}}
\p(974,689){\r[-0.35pt]{1.887pt}{0.7pt}}
\p(982,690){\r[-0.35pt]{1.887pt}{0.7pt}}
\p(990,691){\r[-0.35pt]{1.887pt}{0.7pt}}
\p(998,692){\r[-0.35pt]{1.887pt}{0.7pt}}
\p(1006,693){\r[-0.35pt]{1.887pt}{0.7pt}}
\p(1013,694){\r[-0.35pt]{2.264pt}{0.7pt}}
\p(1023,695){\r[-0.35pt]{2.264pt}{0.7pt}}
\p(1032,696){\r[-0.35pt]{2.264pt}{0.7pt}}
\p(1042,697){\r[-0.35pt]{2.264pt}{0.7pt}}
\p(1051,698){\r[-0.35pt]{2.264pt}{0.7pt}}
\p(1061,699){\r[-0.35pt]{2.831pt}{0.7pt}}
\p(1072,700){\r[-0.35pt]{2.831pt}{0.7pt}}
\p(1084,701){\r[-0.35pt]{2.831pt}{0.7pt}}
\p(1096,702){\r[-0.35pt]{2.831pt}{0.7pt}}
\p(1108,703){\r[-0.35pt]{2.831pt}{0.7pt}}
\p(1119,704){\r[-0.35pt]{2.831pt}{0.7pt}}
\p(1131,705){\r[-0.35pt]{2.831pt}{0.7pt}}
\p(1143,706){\r[-0.35pt]{2.831pt}{0.7pt}}
\p(1155,707){\r[-0.35pt]{2.831pt}{0.7pt}}
\p(1166,708){\r[-0.35pt]{2.831pt}{0.7pt}}
\p(1178,709){\r[-0.35pt]{2.831pt}{0.7pt}}
\p(1190,710){\r[-0.35pt]{2.831pt}{0.7pt}}
\p(1202,711){\r[-0.35pt]{3.694pt}{0.7pt}}
\p(1217,712){\r[-0.35pt]{3.694pt}{0.7pt}}
\p(1232,713){\r[-0.35pt]{3.694pt}{0.7pt}}
\p(1248,714){\r[-0.35pt]{3.774pt}{0.7pt}}
\p(1263,715){\r[-0.35pt]{3.774pt}{0.7pt}}
\p(1279,716){\r[-0.35pt]{3.774pt}{0.7pt}}
\p(1294,717){\r[-0.35pt]{3.774pt}{0.7pt}}
\p(1310,718){\r[-0.35pt]{3.774pt}{0.7pt}}
\p(1326,719){\r[-0.35pt]{3.774pt}{0.7pt}}
\p(1341,720){\r[-0.35pt]{5.661pt}{0.7pt}}
\p(1365,721){\r[-0.35pt]{5.661pt}{0.7pt}}
\p(1389,722){\r[-0.35pt]{5.661pt}{0.7pt}}
\p(1412,723){\r[-0.35pt]{5.661pt}{0.7pt}}
\end{picture}
\begin{center}
\vskip 3.0cm
Fig. 1.
\end{center}
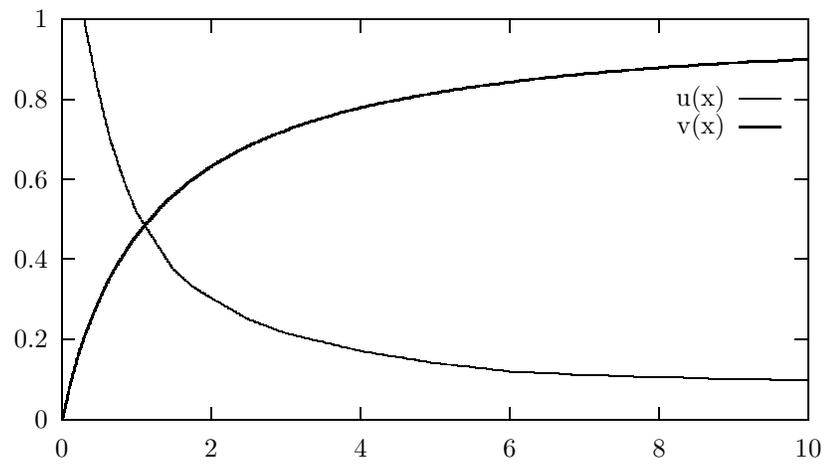
\end{document}